\documentclass[twocolumn]{aastex63}

\usepackage{aas_macros}

\usepackage{color}

\usepackage{natbib}

\usepackage{amsmath}
\usepackage{xspace}
\usepackage{xcolor}
\usepackage{lineno}

\mathchardef\mhyphen="2D

\newlength{\dhatheight}

\providecommand\physrep{\ref@jnl{Phys.~Rep.}}%
\providecommand\apjs{\ref@jnl{ApJS}}%
\providecommand{\jcap}{\ref@jnl{JCAP}}%

\newcommand{\feh}         {\mbox{[Fe/H]}}

\newcommand{\rproc}       {$r$-process}

\def\spose#1{\hbox to 0pt{#1\hss}}
\def\lta{\mathrel{\spose{\lower 3pt\hbox{$\mathchar"218$}}
     \raise 2.0pt\hbox{$\mathchar"13C$}}}
\def\gta{\mathrel{\spose{\lower 3pt\hbox{$\mathchar"218$}}
    \raise 2.0pt\hbox{$\mathchar"13E$}}}

\shorttitle{Timing the $r$-Process Enrichment of Reticulum~II}
\shortauthors{Simon et al.}

\begin{document}

\title{Timing the $r$-Process Enrichment of the Ultra-Faint Dwarf
  Galaxy Reticulum~II}


\author{Joshua~D.~Simon}
\affiliation{Observatories of the Carnegie Institution for Science, 813 Santa Barbara St., Pasadena, CA 91101, USA}
\author{Thomas~M.~Brown}
\affiliation{Space Telescope Science Institute, 3700 San Martin
  Drive, Baltimore, MD  21218, USA}
\author{Bur\c{c}{\rlap{\.}\i}n~Mutlu-Pakd{\rlap{\.}\i}l}
\affiliation{Kavli Institute for Cosmological Physics, University of Chicago, Chicago, IL 60637, USA}
\affiliation{Department of Astronomy and Astrophysics, University of
  Chicago, Chicago, IL 60637, USA}
\affiliation{Department of Physics and Astronomy, Dartmouth College, 6127 Wilder Laboratory, Hanover, NH 03755, USA}
\author{Alexander~P.~Ji}
\affiliation{Department of Astronomy and Astrophysics, University of
  Chicago, Chicago, IL 60637, USA}
\affiliation{Kavli Institute for Cosmological Physics, University of Chicago, Chicago, IL 60637, USA}
\author{Alex~Drlica-Wagner}
\affiliation{Fermi National Accelerator Laboratory, P.O. Box 500,
  Batavia, IL 60510, USA}
\affiliation{Kavli Institute for Cosmological Physics, University of Chicago, Chicago, IL 60637, USA}
\affiliation{Department of Astronomy and Astrophysics, University of
  Chicago, Chicago, IL 60637, USA}
\author{Roberto~J.~Avila}
\affiliation{Space Telescope Science Institute, 3700 San Martin
  Drive, Baltimore, MD  21218, USA}
\author{Clara~E.~Mart{\'i}nez-V{\'a}zquez}
\affiliation{Gemini Observatory, NSF's NOIRLaboratory, 670 N. A'ohoku
  Place, Hilo, HI 96720, USA}
\author{Ting~S.~Li}
\affiliation{Department of Astronomy and Astrophysics, University of
  Toronto, 50 St. George Street, Toronto ON, M5S 3H4, Canada}
\author{Eduardo~Balbinot}
\affiliation{Kapteyn Astronomical Institute, University of Groningen,
  Landleven 12, 9747 AD Groningen, The Netherlands}
\author{Keith~Bechtol}
\affiliation{Department of Physics, University of Wisconsin-Madison,
  Madison, WI 53706, USA}
\author{Anna~Frebel}
\affiliation{Department of Physics \& Kavli Institute for Astrophysics
  and Space Research, Massachusetts Institute of Technology,
  Cambridge, MA 02139, USA}
\affiliation{Joint Institute for Nuclear Astrophysics—Center for
  Evolution of the Elements, East Lansing, MI 48824, USA}
\author{Marla~Geha}
\affiliation{Department of Astronomy, Yale University, 52 Hillhouse
  Avenue, New Haven, CT 06520, USA}
\author{Terese~T.~Hansen}
\affiliation{Department of Astronomy, Stockholm University, AlbaNova
  University Centre, SE-106 91 Stockholm, Sweden}
\author{David~J.~James}
\affiliation{Center for Astrophysics $\vert$ Harvard \& Smithsonian, 60 Garden Street, Cambridge, MA 02138, USA}
\author{Andrew~B.~Pace}
\affiliation{McWilliams Center for Cosmology, Carnegie Mellon
  University, 5000 Forbes Avenue, Pittsburgh, PA 15213, USA}
\author{M.~Aguena}
\affiliation{Laborat\'orio Interinstitucional de e-Astronomia - LIneA, Rua Gal. Jos\'e Cristino 77, Rio de Janeiro, RJ - 20921-400, Brazil}
\author{O.~Alves}
\affiliation{Department of Physics, University of Michigan, Ann Arbor, MI 48109, USA}
\author{F.~Andrade-Oliveira}
\affiliation{Department of Physics, University of Michigan, Ann Arbor, MI 48109, USA}
\author{J.~Annis}
\affiliation{Fermi National Accelerator Laboratory, P. O. Box 500, Batavia, IL 60510, USA}
\author{D.~Bacon}
\affiliation{Institute of Cosmology and Gravitation, University of Portsmouth, Portsmouth, PO1 3FX, UK}
\author{E.~Bertin}
\affiliation{CNRS, UMR 7095, Institut d'Astrophysique de Paris, F-75014, Paris, France}
\affiliation{Sorbonne Universit\'es, UPMC Univ Paris 06, UMR 7095, Institut d'Astrophysique de Paris, F-75014, Paris, France}
\author{D.~Brooks}
\affiliation{Department of Physics \& Astronomy, University College London, Gower Street, London, WC1E 6BT, UK}
\author{D.~L.~Burke}
\affiliation{Kavli Institute for Particle Astrophysics \& Cosmology, P. O. Box 2450, Stanford University, Stanford, CA 94305, USA}
\affiliation{SLAC National Accelerator Laboratory, Menlo Park, CA 94025, USA}
\author{A.~Carnero~Rosell}
\affiliation{Instituto de Astrofisica de Canarias, E-38205 La Laguna, Tenerife, Spain}
\affiliation{Laborat\'orio Interinstitucional de e-Astronomia - LIneA, Rua Gal. Jos\'e Cristino 77, Rio de Janeiro, RJ - 20921-400, Brazil}
\affiliation{Universidad de La Laguna, Dpto. Astrofísica, E-38206 La Laguna, Tenerife, Spain}
\author{M.~Carrasco~Kind}
\affiliation{Center for Astrophysical Surveys, National Center for Supercomputing Applications, 1205 West Clark St., Urbana, IL 61801, USA}
\affiliation{Department of Astronomy, University of Illinois at Urbana-Champaign, 1002 W. Green Street, Urbana, IL 61801, USA}
\author{J.~Carretero}
\affiliation{Institut de F\'{\i}sica d'Altes Energies (IFAE), The Barcelona Institute of Science and Technology, Campus UAB, 08193 Bellaterra (Barcelona) Spain}
\author{M.~Costanzi}
\affiliation{Astronomy Unit, Department of Physics, University of Trieste, via Tiepolo 11, I-34131 Trieste, Italy}
\affiliation{INAF-Osservatorio Astronomico di Trieste, via G. B. Tiepolo 11, I-34143 Trieste, Italy}
\affiliation{Institute for Fundamental Physics of the Universe, Via Beirut 2, 34014 Trieste, Italy}
\author{L.~N.~da Costa}
\affiliation{Laborat\'orio Interinstitucional de e-Astronomia - LIneA, Rua Gal. Jos\'e Cristino 77, Rio de Janeiro, RJ - 20921-400, Brazil}
\author{J.~De~Vicente}
\affiliation{Centro de Investigaciones Energ\'eticas, Medioambientales y Tecnol\'ogicas (CIEMAT), Madrid, Spain}
\author{S.~Desai}
\affiliation{Department of Physics, IIT Hyderabad, Kandi, Telangana 502285, India}
\author{P.~Doel}
\affiliation{Department of Physics \& Astronomy, University College London, Gower Street, London, WC1E 6BT, UK}
\author{S.~Everett}
\affiliation{Jet Propulsion Laboratory, California Institute of Technology, 4800 Oak Grove Dr., Pasadena, CA 91109, USA}
\author{I.~Ferrero}
\affiliation{Institute of Theoretical Astrophysics, University of Oslo. P.O. Box 1029 Blindern, NO-0315 Oslo, Norway}
\author{J.~Frieman}
\affiliation{Fermi National Accelerator Laboratory, P. O. Box 500, Batavia, IL 60510, USA}
\affiliation{Kavli Institute for Cosmological Physics, University of Chicago, Chicago, IL 60637, USA}
\author{J.~Garc\'ia-Bellido}
\affiliation{Instituto de Fisica Teorica UAM/CSIC, Universidad Autonoma de Madrid, 28049 Madrid, Spain}
\author{M.~Gatti}
\affiliation{Department of Physics and Astronomy, University of Pennsylvania, Philadelphia, PA 19104, USA}
\author{D.~W.~Gerdes}
\affiliation{Department of Astronomy, University of Michigan, Ann Arbor, MI 48109, USA}
\affiliation{Department of Physics, University of Michigan, Ann Arbor, MI 48109, USA}
\author{D.~Gruen}
\affiliation{University Observatory, Faculty of Physics, Ludwig-Maximilians-Universit\"at, Scheinerstr. 1, 81679 Munich, Germany}
\author{R.~A.~Gruendl}
\affiliation{Center for Astrophysical Surveys, National Center for Supercomputing Applications, 1205 West Clark St., Urbana, IL 61801, USA}
\affiliation{Department of Astronomy, University of Illinois at Urbana-Champaign, 1002 W. Green Street, Urbana, IL 61801, USA}
\author{J.~Gschwend}
\affiliation{Laborat\'orio Interinstitucional de e-Astronomia - LIneA, Rua Gal. Jos\'e Cristino 77, Rio de Janeiro, RJ - 20921-400, Brazil}
\affiliation{Observat\'orio Nacional, Rua Gal. Jos\'e Cristino 77, Rio de Janeiro, RJ - 20921-400, Brazil}
\author{G.~Gutierrez}
\affiliation{Fermi National Accelerator Laboratory, P. O. Box 500, Batavia, IL 60510, USA}
\author{S.~R.~Hinton}
\affiliation{School of Mathematics and Physics, University of Queensland,  Brisbane, QLD 4072, Australia}
\author{D.~L.~Hollowood}
\affiliation{Santa Cruz Institute for Particle Physics, Santa Cruz, CA 95064, USA}
\author{K.~Honscheid}
\affiliation{Center for Cosmology and Astro-Particle Physics, The Ohio State University, Columbus, OH 43210, USA}
\affiliation{Department of Physics, The Ohio State University, Columbus, OH 43210, USA}
\author{K.~Kuehn}
\affiliation{Australian Astronomical Optics, Macquarie University, North Ryde, NSW 2113, Australia}
\affiliation{Lowell Observatory, 1400 Mars Hill Rd, Flagstaff, AZ 86001, USA}
\author{N.~Kuropatkin}
\affiliation{Fermi National Accelerator Laboratory, P. O. Box 500, Batavia, IL 60510, USA}
\author{J.~L.~Marshall}
\affiliation{George P. and Cynthia Woods Mitchell Institute for Fundamental Physics and Astronomy, and Department of Physics and Astronomy, Texas A\&M University, College Station, TX 77843,  USA}
\author{J. Mena-Fern{\'a}ndez}
\affiliation{Centro de Investigaciones Energ\'eticas, Medioambientales y Tecnol\'ogicas (CIEMAT), Madrid, Spain}
\author{R.~Miquel}
\affiliation{Instituci\'o Catalana de Recerca i Estudis Avan\c{c}ats, E-08010 Barcelona, Spain}
\affiliation{Institut de F\'{\i}sica d'Altes Energies (IFAE), The Barcelona Institute of Science and Technology, Campus UAB, 08193 Bellaterra (Barcelona) Spain}
\author{A.~Palmese}
\affiliation{Department of Astronomy, University of California, Berkeley,  501 Campbell Hall, Berkeley, CA 94720, USA}
\author{F.~Paz-Chinch\'{o}n}
\affiliation{Center for Astrophysical Surveys, National Center for Supercomputing Applications, 1205 West Clark St., Urbana, IL 61801, USA}
\affiliation{Institute of Astronomy, University of Cambridge, Madingley Road, Cambridge CB3 0HA, UK}
\author{M.~E.~S.~Pereira}
\affiliation{Hamburger Sternwarte, Universit\"{a}t Hamburg, Gojenbergsweg 112, 21029 Hamburg, Germany}
\author{A.~Pieres}
\affiliation{Laborat\'orio Interinstitucional de e-Astronomia - LIneA, Rua Gal. Jos\'e Cristino 77, Rio de Janeiro, RJ - 20921-400, Brazil}
\affiliation{Observat\'orio Nacional, Rua Gal. Jos\'e Cristino 77, Rio de Janeiro, RJ - 20921-400, Brazil}
\author{A.~A.~Plazas~Malag\'on}
\affiliation{Department of Astrophysical Sciences, Princeton University, Peyton Hall, Princeton, NJ 08544, USA}
\author{M.~Raveri}
\affiliation{Department of Physics and Astronomy, University of Pennsylvania, Philadelphia, PA 19104, USA}
\author{M.~Rodriguez-Monroy}
\affiliation{Centro de Investigaciones Energ\'eticas, Medioambientales y Tecnol\'ogicas (CIEMAT), Madrid, Spain}
\author{E.~Sanchez}
\affiliation{Centro de Investigaciones Energ\'eticas, Medioambientales y Tecnol\'ogicas (CIEMAT), Madrid, Spain}
\author{B.~Santiago}
\affiliation{Instituto de F\'\i sica, UFRGS, Caixa Postal 15051, Porto Alegre, RS - 91501-970, Brazil}
\affiliation{Laborat\'orio Interinstitucional de e-Astronomia - LIneA, Rua Gal. Jos\'e Cristino 77, Rio de Janeiro, RJ - 20921-400, Brazil}
\author{V.~Scarpine}
\affiliation{Fermi National Accelerator Laboratory, P. O. Box 500, Batavia, IL 60510, USA}
\author{I.~Sevilla-Noarbe}
\affiliation{Centro de Investigaciones Energ\'eticas, Medioambientales y Tecnol\'ogicas (CIEMAT), Madrid, Spain}
\author{M.~Smith}
\affiliation{School of Physics and Astronomy, University of Southampton,  Southampton, SO17 1BJ, UK}
\author{E.~Suchyta}
\affiliation{Computer Science and Mathematics Division, Oak Ridge National Laboratory, Oak Ridge, TN 37831}
\author{M.~E.~C.~Swanson}
\affiliation{}
\author{G.~Tarle}
\affiliation{Department of Physics, University of Michigan, Ann Arbor, MI 48109, USA}
\author{C.~To}
\affiliation{Center for Cosmology and Astro-Particle Physics, The Ohio State University, Columbus, OH 43210, USA}
\author{M.~Vincenzi}
\affiliation{Institute of Cosmology and Gravitation, University of Portsmouth, Portsmouth, PO1 3FX, UK}
\affiliation{School of Physics and Astronomy, University of Southampton,  Southampton, SO17 1BJ, UK}
\author{N.~Weaverdyck}
\affiliation{Department of Physics, University of Michigan, Ann Arbor, MI 48109, USA}
\affiliation{Lawrence Berkeley National Laboratory, 1 Cyclotron Road, Berkeley, CA 94720, USA}
\author{R.~D.~Wilkinson}
\affiliation{Department of Physics and Astronomy, Pevensey Building, University of Sussex, Brighton, BN1 9QH, UK}

\collaboration{68}{(DES Collaboration)}

\begin{abstract}
  The ultra-faint dwarf galaxy Reticulum~II (Ret~II) exhibits a unique
  chemical evolution history, with $72^{+10}_{-12}$\% of its stars
  strongly enhanced in \rproc\ elements.  We present deep Hubble Space
  Telescope photometry of Ret~II and analyze its star formation
  history.  As in other ultra-faint dwarfs, the color-magnitude
  diagram is best fit by a model consisting of two bursts of star
  formation.  If we assume that the bursts were instantaneous, then
  the older burst occurred around the epoch of reionization and formed
  $\sim80$\%\ of the stars in the galaxy, while the remainder of the
  stars formed $\sim3$~Gyr later.  When the bursts are allowed to have
  nonzero durations we obtain slightly better fits.  The best-fitting
  model in this case consists of two bursts beginning before
  reionization, with approximately half the stars formed in a short
  (100~Myr) burst and the other half in a more extended period lasting
  2.6~Gyr.  Considering the full set of viable star formation history
  models, we find that 28\%\ of the stars formed within $500 \pm
  200$~Myr of the onset of star formation.  The combination of the
  star formation history and the prevalence of \rproc-enhanced stars
  demonstrates that the \rproc\ elements in Ret~II must have been
  synthesized early in its initial star-forming phase.  We therefore
  constrain the delay time between the formation of the first stars in
  Ret~II and the \rproc\ nucleosynthesis to be less than 500~Myr.
  This measurement rules out an \rproc\ source with a delay time of
  several Gyr or more such as GW170817.
\end{abstract}


\section{INTRODUCTION}
\label{intro}

Identifying the astrophysical production site of the \rproc\ elements
has been a long-standing goal of studies of chemical evolution and
nuclear astrophysics \citep[e.g.,][and references therein]{frebel18}.
At the broadest level, the debate is between rare events producing
large quantities of \rproc\ material, or frequent events producing
small amounts of \rproc\ elements.  The former category includes
neutron star mergers \citep[e.g.,][]{ls74,meyer89} and jet-driven
supernovae \citep[e.g.,][]{cameron03,fujimoto08,hm18}, and more
recently, collapsars \citep[e.g.,][]{mw99,surman06,siegel19}, while
the latter is usually assumed to relate to ordinary core-collapse
supernova (SN) explosions \citep[e.g.,][]{b2fh,at13}.

The discovery of the strongly \rproc-enhanced stars ($\textrm{[Eu/Fe]}
> 1.0$) in the ultra-faint dwarf (UFD) galaxy Reticulum~II (Ret~II)
provided decisive evidence in favor of a rare and prolific source of
the \rproc\ elements \citep{ji16,roederer16}.  The subsequent
confirmation of \rproc\ nucleosynthesis in the neutron star merger
GW170817 \citep{chornock17,drout17,kasliwal17,pian17,tanvir17} then
made a strong case that the rare and prolific source should be
identified with merging neutron stars.  However, the story is not
necessarily over.  The combination of the short expected time scale
for star formation in Ret~II and the potentially long delay time
between star formation and a neutron star merger is a challenge to
understand.  Specifically, although the core-collapse explosions of
massive stars occur within a few million years of the birth of those
stars, binary neutron star systems may take hundreds of Myr or even
longer to merge \citep[e.g.,][]{dominik12}.  The relative importance
of neutron star mergers and rapid sources such as collapsars and
jet-driven supernovae to \rproc\ nucleosynthesis also remains
controversial
\citep[e.g.,][]{siegel19,bartos19,brauer21,reggiani21,fs22}.  And the
low but non-zero levels of \rproc\ enrichment in other UFDs appear to
require a second source (presumably core-collapse SNe) of
\rproc\ material as well
\citep[e.g.,][]{frebel10,fsk14,ishigaki14,ji16c}.

Dwarf galaxies provide clean environments for unraveling early
galactic chemical evolution because of the limited number of
nucleosynthetic events that enriched their oldest stars
\citep[e.g.,][]{simon15b,cw19,hartwig19} and because each dwarf is the
descendant of relatively few progenitor systems
\citep{fitts18,griffen18}.  Ret~II is particularly interesting in this
regard, as $72^{+10}_{-12}$\% of its stars are \rproc-enhanced
\citep[][]{ji16,ji16b,ji22}.  The non-\rproc-enhanced fraction is thus
$28^{+12}_{-10}$\%.  This result demonstrates that the event that
produced the \rproc\ elements in Ret~II must have occurred at a time
when no more than 28\% of the galaxy's stars had formed.
Given the contrast in time scales between different candidate
\rproc\ sites, determining the star formation history (SFH) of Ret~II
therefore offers the possibility of placing a limit on the time scale
of \rproc-enrichment in the galaxy, which could enable a prompt or
delayed source for the \rproc\ elements to be distinguished.

The stellar populations of the UFDs that have been examined thus far
are exclusively old, with typical ages of $\sim13$~Gyr
\citep{brown12,brown14}.  However, there is tentative evidence for
some differences in detail from galaxy to galaxy.  For example,
Ursa~Major~I may exhibit a longer duration of star formation and
younger mean age than the other UFDs in the \citet{brown14} sample,
most notably Canes~Venatici~II (CVn~II) and Coma~Berenices (Com),
although to a lesser extent Bo{\" o}tes~I, Hercules, and Leo~IV as
well.  Intriguingly, \citet{sacchi21} also detected a possible
difference in mean age between satellites of the Magellanic Clouds
(including Ret~II) and satellites of the Milky Way, with star
formation in the former systems ending 600~Myr later.

The \citet{sacchi21} SFH for Ret~II is based on imaging of a single
Hubble Space Telescope (HST) pointing that covers a small fraction of
the galaxy and therefore contains less than 200 Ret~II stars
($\sim2$\% of the total stellar content, assuming the stellar mass of
3300~M$_{\odot}$ from \citealt{ji22}, and the initial mass function
from \citealt{safarzadeh21}).  In this paper, we present a SFH
analysis of Ret~II from wider-area HST imaging with a substantially
larger sample of stars ($\sim2600$~Ret~II members; $\sim25$\% of the
stars in the galaxy).  In \S~\ref{sec:data}, we describe our
observations and the data reduction and photometry procedures.  In
\S~\ref{sec:sfh}, we determine the SFH of Ret~II.  We discuss the
implications of the SFH for \rproc\ nucleosynthesis in
\S~\ref{sec:discussion}, and we summarize our results and conclusions
in \S~\ref{sec:conclusion}.

\section{HUBBLE SPACE TELESCOPE DATA}
\label{sec:data}

\subsection{Observations}
\label{sec:obs}

We obtained a 12-tile mosaic of Ret~II with the Wide Field Channel of
the Advanced Camera for Surveys (ACS; \citealt{acs}) on HST through
program GO-14766 (PI: Simon).  The observations were carried out
between 2016 November 9 and 2016 November 26.

Given the proximity of Ret~II to the Milky Way, even short integration
times enabled us to exceed the target signal-to-noise ratio (S/N) of
100 at 1~mag below the oldest main sequence turnoff (MSTO), which
occurs at $m_{814} \approx 23.5$ at the distance of Ret~II.  Each
mosaic tile was observed for a single orbit, consisting of two F606W
exposures and two F814W exposures, with total integration times of
980~s in F606W and 1140~s in F814W.  A 3\farcs034 dither was performed
between the first and second exposures in each filter to fill in the
gap between the ACS chips and reject cosmic rays.  In previous UFD
programs with HST \citep{brown14,simon21}, we used a four-point dither
pattern to fully sample the ACS point-spread function (PSF).  However,
CCD readout constraints make it impossible to obtain eight exposures
across multiple filters in one HST orbit, so additional dither
positions would have required doubling the number of orbits devoted to
the project or halving the area covered.  Being limited to two
dithered exposures per band has implications for the photometric
methods employed, as discussed below, but was workable given the low
surface density of stars in Ret~II.

Based on the distribution of Ret~II member stars in the initial Dark
Energy Survey data \citep{bechtol15,dw18}, the mosaic tiles were
arranged in an elongated east-west pattern to completely cover the
area within the half-light radius of Ret~II.  The main east-west
mosaic pattern included only one blue horizontal branch (BHB) star
from the spectroscopic sample of \citet{simon15}, so one ACS tile was
placed in the northeast corner of the mosaic to observe a second BHB
star in order to better constrain the distance of Ret~II.  The spatial
coverage of the observations is shown in Figure~1 of
\citet{safarzadeh21}.

\subsection{Reduction and Photometry}
\label{sec:reduction}
Our initial data reduction procedures followed those described by
\citet{brown14} and \citet{simon21}.  We processed the raw images with
the most recent version of the ACS pipeline, including bias
subtraction, dark subtraction, identification of detector artifacts,
and charge transfer inefficiency correction.  The images were
resampled onto a 0\farcs035 pixel grid, with each tile covering an
area of approximately $200\arcsec \times 205\arcsec$, and overlaps of
a few arcsec between adjacent tiles.

\begin{figure*}
\epsscale{1.2}
\plotone{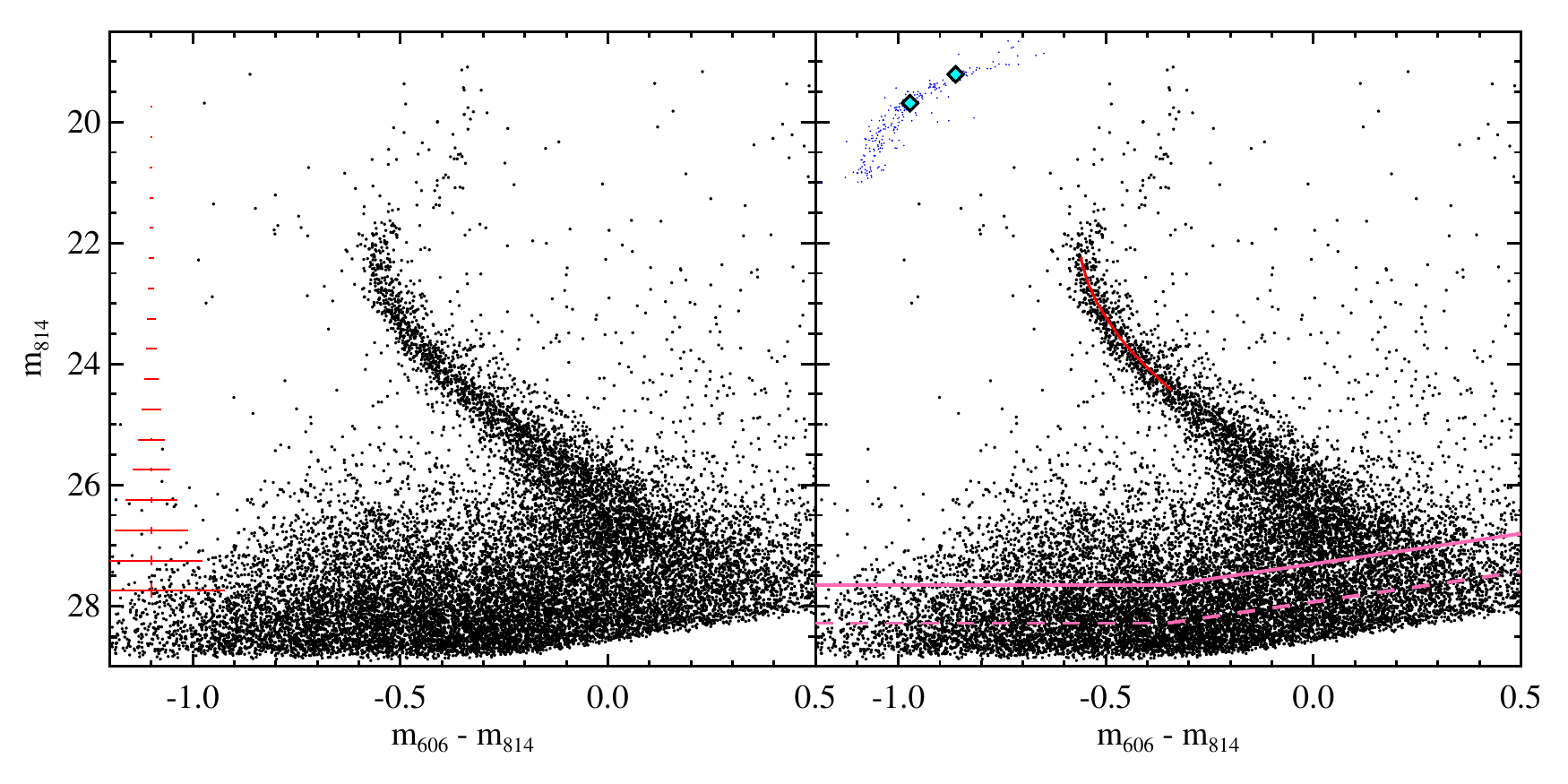}
\caption{(Left) ACS color-magnitude diagram of Ret~II.  Magnitude
  measurements are in the STMAG system, and only stars meeting the
  criteria described in \S~\ref{sec:reduction} are included.  The two
  stars closest to the upper left corner are the BHB members of
  Ret~II.  Typical photometric uncertainties as a function of
  magnitude are plotted in red along the left of the
  figure. (Right) Ret~II color-magnitude diagram, as in the left
  panel, with the M92 horizontal branch (blue dots in upper left) and
  the upper main sequence portion of the Victoria-Regina isochrone
  (red curve) overlaid, both shifted to the best-fit distance and
  reddening for Ret~II.  The Ret~II BHB stars are highlighted with
  cyan diamonds.  The 90\% and 50\% completeness limits are displayed
  as solid and dashed pink lines, respectively.}
\label{fig:cmd}
\end{figure*}

We performed PSF photometry on the pipeline-produced flat-fielded,
charge transfer efficiency-corrected (FLC) images with the latest
version (2.0) of DOLPHOT \citep{dolphin00}, as in \citet{mp19,mp20}.
We followed the recommended preprocessing steps and used the suggested
input parameters from the DOLPHOT User
Guide\footnote{\url{http://americano.dolphinsim.com/dolphot/dolphotACS.pdf}}.
The initial photometry was then cleaned of spurious detections using
the following criteria: the sum of the crowding parameters in the two
bands must be $<1$, the squared sum of the sharpness parameters in the
two bands must be $<0.1$, and the signal-to-noise ratio must be
$\geq4$ and object-type\footnote{This parameter distinguishes point
  sources from extended sources and artifacts, as described in the
  DOLPHOT manual.} must be $\leq2$ in each band.  The resulting CMD is
displayed in Fig.~\ref{fig:cmd}, and Table~\ref{photom_table} contains
the cleaned photometric catalog.

\begin{deluxetable*}{llllllr}
\tablecaption{Ret~II Stellar Photometry}
\tablewidth{0pt}
\tablehead{
\colhead{Star} & \colhead{RA (J2000)} & \colhead{Dec J2000)} & \colhead{$m_{606}$}
& \colhead{$\delta m_{606}$} & \colhead{$m_{814}$} & \colhead{$\delta m_{814}$}
}
\startdata
    1 &  54.123158 & $-$53.961302 & 20.233 & 0.002 & 19.410 & 0.002 \\
    2 &  54.108976 & $-$53.968195 & 20.201 & 0.002 & 20.081 & 0.002 \\
    3 &  54.132862 & $-$53.957064 & 21.105 & 0.003 & 19.781 & 0.001 \\
    4 &  54.102355 & $-$53.948340 & 20.669 & 0.002 & 20.271 & 0.002 \\
    5 &  54.086859 & $-$53.943280 & 21.470 & 0.003 & 20.241 & 0.002 \\
    6 &  54.123401 & $-$53.946231 & 20.995 & 0.003 & 21.415 & 0.003 \\
    7 &  54.149917 & $-$53.941306 & 21.129 & 0.003 & 21.503 & 0.003 \\
    8 &  54.143632 & $-$53.958409 & 22.183 & 0.005 & 20.967 & 0.003 \\
    9 &  54.134904 & $-$53.966469 & 22.181 & 0.005 & 21.030 & 0.002 \\
   10 &  54.059216 & $-$53.967990 & 21.897 & 0.004 & 21.208 & 0.003 \\
\enddata
\tablecomments{Magnitudes are on the STMAG system, and astrometry is calibrated to Gaia~DR3 \citep{gaiadr3vallenari}.  This table is available in its entirety in machine-readable form.}
\label{photom_table}
\end{deluxetable*}

We carried out artificial star tests in order to quantify the
photometric uncertainties and completeness in our observations, using
the artificial star utilities in DOLPHOT. We injected a total of
$\sim$5 million artificial stars per tile, distributing them evenly
across the field of view.  Because each star is inserted and
photometered one at a time, the large number of stars inserted during
the artificial star tests does not cause any self-induced crowding
\citep{dolphin00}. The input colors and magnitudes of the artificial
stars covered the complete range of the observed colors and magnitudes
(i.e., $18 \leq m_{606} \leq 30$ and $-0.75\leq m_{606}-m_{814}
\leq2.0$).  Photometry and quality cuts were performed in an identical
manner to those performed on the original photometry.

The two BHB stars mentioned in \S~\ref{sec:obs} were saturated in the
drizzled images.  Because these stars are important for determining
the distance of Ret~II (\S~\ref{sec:distance}), we used aperture
photometry on the individual FLC exposures to recover their fluxes by
hand.  The instrumental magnitudes measured in this way from the two
separate exposures per band in each tile agreed to better than
0.01~mag for each star, indicating that the photometric precision for
these stars remains high despite the saturation effects.

We note that in past SFH analyses by our group
\citep[e.g.,][]{brown14,simon21}, we have relied on PSF-fitting
photometry with the DAOPHOT-II package \citep{stetson87}.  With this
data set, applying the same procedures we had used in the past
resulted in unexpected artifacts in the color-magnitude diagram (CMD),
most notably an increased width of the main sequence, which were not
present in photometry carried out with DOLPHOT.  The difference
between the DAOPHOT and DOLPHOT results appears to be a consequence of
the incomplete sampling of the PSF from having only two dither
positions and the sparseness of the field.  In this regime, the
empirical PSF library used by DOLPHOT is more appropriate than a PSF
model constructed directly from the data in DAOPHOT.

\section{THE STAR FORMATION HISTORY OF RET~II}
\label{sec:sfh}

\subsection{Metallicity Distribution, Distance, and Reddening}
\label{sec:distance}

Before modeling the SFH of Ret~II, we first determined the distance of
the galaxy, the foreground reddening, and the metallicity distribution
in order to be able to compare theoretical stellar isochrones with the
HST photometry.

Distances to Local Group dwarf galaxies are typically best determined
with RR~Lyrae variable stars
\citep[e.g.,][]{martinezvazquez17,martinezvazquez19,hernitschek19,muraveva20,nagarajan22}.
Unfortunately, no RR~Lyrae have been identified in Ret~II
\citep{vivas20}.  The other features in the color-magnitude diagram
for an old, metal-poor stellar population that are good distance
indicators are the main sequence turnoff and the horizontal branch.
Following \citet{brown14} and \citet{simon21}, we simultaneously fit
the main sequence of Ret~II with a Victoria-Regina isochrone
\citep{vdb14} and the two BHB stars with the horizontal branch of the
old, metal-poor globular cluster M92 (see Fig.~\ref{fig:cmd}).  We
assumed a distance modulus of 14.62~mag
\citep{delprincipe05,sollima06,paust07} and foreground reddening of
$E(B-V) = 0.023$~mag \citep{sfd98} for M92.  For the theoretical
isochrone, we assumed an age of 13 Gyr and the metallicity
distribution described below, as well as a binary fraction of 0.48
\citep{geha13}.  We found a distance modulus for Ret~II of $m-M =
17.50$~mag ($d = 31.6$~kpc) and reddening of $E(B-V) =
0.052$~mag.\footnote{For comparison, the reddening at the center of
  Ret~II according to \citet{sf11} is $E(B-V) = 0.016$~mag.  The
  reddening determined for the other UFDs analyzed by \citet{brown14}
  and \citet{simon21} with the same methodology was also larger than
  indicated by dust maps.}  As in our previous analyses, we assumed
uncertainties of 0.07~mag in the distance modulus and 0.01~mag in
$E(B-V)$.  The distance of Ret~II is in excellent agreement with the
measurements of \citet{mp18} and \citet{bechtol15}, although our
reddening value is substantially larger.

For the metallicity distribution function (MDF) of Ret~II, we relied
on the spectroscopic sample of 16 stars with Ca triplet-based
metallicities from \citet{simon15}.  The other previously published
studies of Ret~II contain the same or smaller samples of member stars
\citep{koposov15,ji16,roederer16}.  Under the assumption of a Gaussian
MDF, \citet{simon15} determined that the width of the distribution is
$\sigma_{\textrm{[Fe/H]}} = 0.28 \pm 0.09$~dex. The most recent
spectroscopic analysis by \citet{ji22} included a larger set of Ret~II
members, but their Fe abundances are based on a single line, and many
of the stars have only upper limits on [Fe/H].  Nevertheless, their
Gaussian MDF agreed with that of \citet{simon15}, finding
$\sigma_{\textrm{[Fe/H]}} = 0.32^{+0.10}_{-0.07}$~dex.

\begin{figure}
\epsscale{1.2}
\plotone{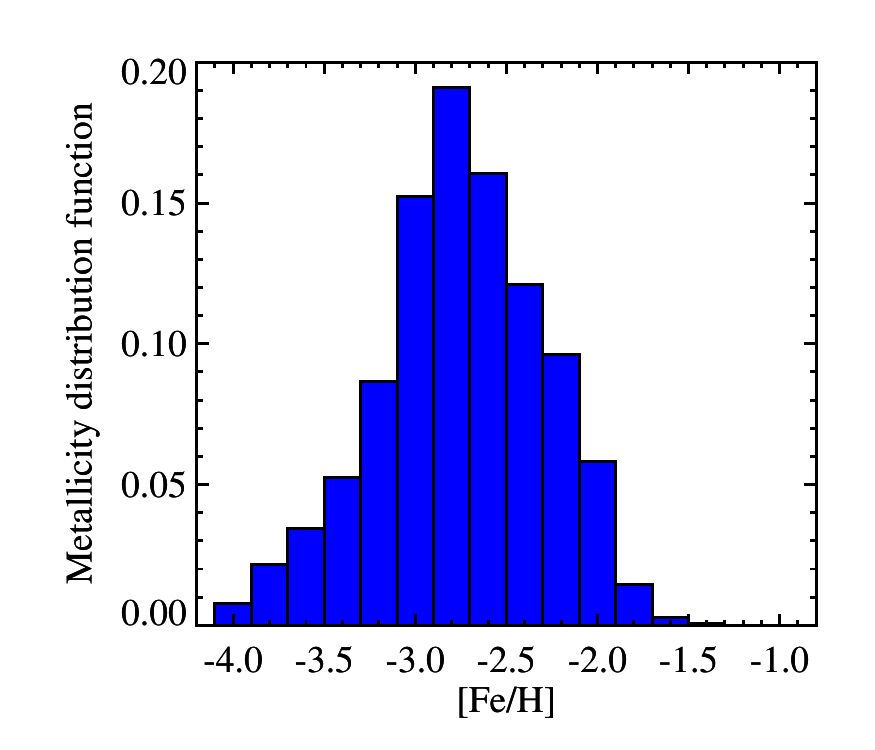}
\caption{Metallicity distribution of Ret~II, as determined from
    the spectroscopic metallicity measurements of \citet{simon15}.
    Each individual stellar metallicity is modeled as a Gaussian
    probability distribution, and samples are drawn from the combined
    probability distribution from all 16 stars.}
\label{fig:mdf}
\end{figure}

To convert these metallicity measurements into an MDF without the
assumption of an overall Gaussian shape, we modeled the metallicity of
each star as a Gaussian probability distribution, constructed the
cumulative distribution of the full set of metallicities, and then
drew 16 samples from the cumulative distribution.  Repeating this
process $10^{5}$ times, we built up a binned MDF for Ret~II (see
Figure~\ref{fig:mdf}).  As described by \citet{simon21}, this process
produces an MDF that is somewhat broader than the true metallicity
distribution because it convolves the intrinsic MDF of the galaxy with
the observational uncertainties.  Given the observed width of the
Ret~II MDF \citep{simon15}, though, we do not expect this broadening
to have a significant effect.  In addition, we note that although the
MDF constrains the set of isochrones used to model the SFH, the
position of old isochrones in the F606W--F814W CMD is a very weak
function of metallicity at $\feh < -2$.

\subsection{Star Formation History Modeling}
\label{sec:sfh_modeling}

As a starting point, we modeled the SFH of Ret~II using the same
techniques as \citet{brown14} and \citet{simon21}, to which the reader
should refer for more details.  We created a Hess diagram from the ACS
CMD, with bins of 0.02~mag in both color and magnitude.  We used
Besan{\c c}on model simulations \citep{besancon} to evaluate
contamination from foreground Milky Way stars in the ACS photometry.
Given the large area covered by the HST mosaic, the estimated
contamination in the region occupied by Ret~II stars was 3.5\%.  To
determine ages, we built model Hess diagrams based on linear
combinations of Victoria-Regina isochrones, assuming the best-fit
values of the initial mass function and binary fraction determined by
\citet{geha13} for Hercules\footnote{The \citet{safarzadeh21}
  measurements of the initial mass function and binary fraction for
  Ret~II are consistent with the \citet{geha13} results for
  Hercules.}, applying the constraints that the combination must have
a metallicity distribution matching the observed MDF and that the
metallicity increases with time.  The isochrone grid used for the
synthetic Hess diagrams spanned from $\rm{[Fe/H]} = -4$ to
$\rm{[Fe/H]} = -1$ in metallicity and 8--14.5~Gyr in
age.\footnote{Although 14.5~Gyr is nominally older than the age of the
  universe in the standard cosmology, as discussed by \citet{brown14},
  the ages in this paper should be regarded as ages relative to a
  model in which the age of M92 is 13.2~Gyr.\label{age_footnote}} We
then fit the observed Hess diagram with the set of models in a region
around the main sequence turnoff using the Poisson likelihood ratio
from \citet{dolphin02}.

\begin{figure}
\epsscale{1.14}
\plotone{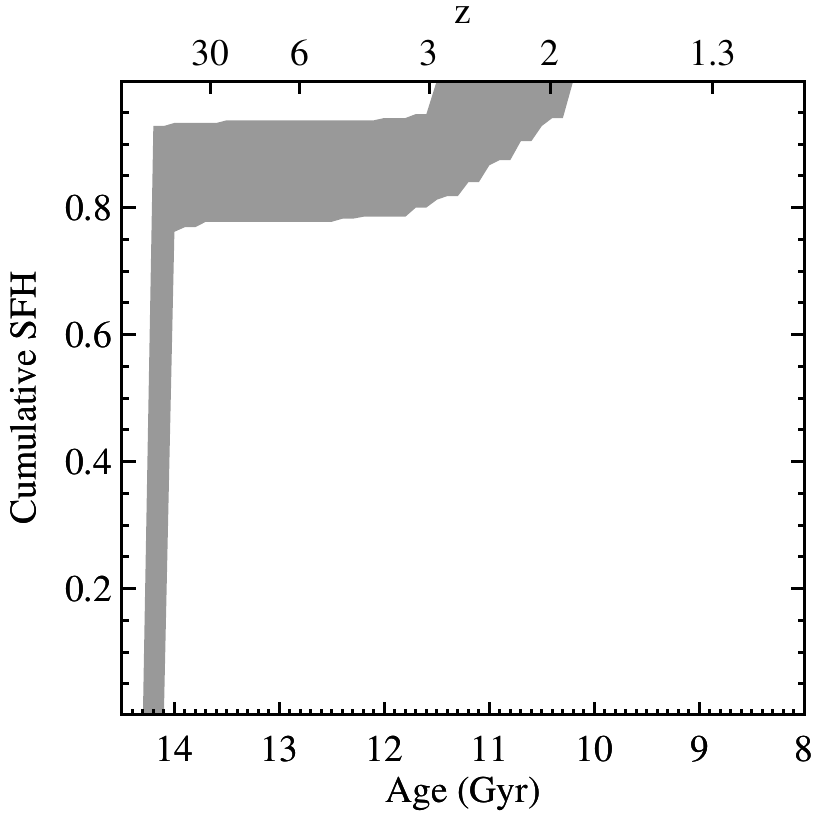}
\caption{Star formation history of Ret~II with a model consisting of
  two instantaneous bursts. This figure can be directly compared to
  the SFHs of other UFDs determined by \citet{brown14}.}
\label{fig:sfh}
\end{figure}

Based on our previous work, our initial model for the Ret~II SFH
consisted of two instantaneous bursts of star formation, with the
timing of the bursts and the fraction of stars formed in each burst as
free parameters.  In this model, the best fit consisted of a burst
that occurred 14.1~Gyr ago containing 87.5\% of the stars and a second
burst at 10.7~Gyr ago with the remaining 12.5\% of the stars.  The
range of SFHs for Ret~II in this model is shown in Fig.~\ref{fig:sfh},
where the shaded region is based on the set of parameters that produce
fit results within $1\sigma$ of the single best fit.  This SFH is
similar to those determined for Canes~Venatici~II and Coma~Berenices
by \citet{brown14}, with $\gtrsim80$\% of the stars formed nearly
immediately after the Big Bang and the possibility of a small fraction
forming up to a few Gyr later.  As described in
footnote~\ref{age_footnote} above, ages older than 14 Gyr for these
UFDs should not be taken as an indication of an inconsistency with the
age of the universe according to current cosmological models, but
simply that star formation in these systems began $\sim1$~Gyr earlier
than in M92.  Our Ret~II SFH also agrees within the uncertainties with
the measurements from the independent HST photometry of
\citet{sacchi21}.

Because our goal in this study was to constrain the timing of the
\rproc\ enrichment of Ret~II, which cannot be accomplished when $>
80$\% of the stars form instantaneously, we also considered models in
which the two bursts\footnote{Although we continue to use the term
  ``burst'' in this context, simulations suggest that such an extended
  star formation episode likely consists of a number of discrete
  bursts interspersed with quiescent periods as a result of stellar
  feedback \citep[e.g.,][]{jeon17,wheeler19}.} have non-zero duration.
\citet{brown14} tried similar models for the six UFDs they analyzed
and found that increasing the burst duration did not improve the fits,
but in the case of Ret~II we obtained a different result.  To
establish reasonable boundaries for the parameter space, motivated by
the results of the instantaneous burst model fit above, we imposed
these conditions: (1) the first burst began between 11.5 and 14.5~Gyr
ago, (2) the duration of the first burst was 0--3~Gyr if it started
11.5--13.5~Gyr ago, or 0--5~Gyr if it started 13.5--14.5~Gyr ago, (3)
the second burst began between the start of the first burst and 8~Gyr
ago, and (4) the second burst can have a duration up to 3 Gyr, but it
must have ended by 8 Gyr ago (e.g., a burst that started at 8.5 Gyr
could only last for 0.5 Gyr).  The limit of no star formation more
recent than 8~Gyr ago was based on the results obtained with the
instantaneous burst model.  With this extended burst model, the best
fit consisted of one burst beginning 14.3~Gyr ago and continuing for
2.6~Gyr, comprising 56\% of the stars, and a second short burst
beginning 14.2~Gyr ago, lasting 100~Myr, and containing 44\% of the
stars.  We emphasize, though, that many other combinations of two
bursts are also consistent with the data, including those with both
burst durations significantly exceeding 100~Myr.  Unlike the modeling
of other UFDs, for Ret~II the extended bursts produced a SFH with a
maximum likelihood score that was $1.4\sigma$ better than that
achieved by the instantaneous burst model.  Because the extended burst
model used five free parameters, whereas the instantaneous burst model
had three free parameters, it is worth noting that the extended burst
fit is superior even if one penalizes for the number of free
parameters, using either the Bayesian Information Criterion or the
Akaike Information Criterion.

\begin{figure}
\epsscale{1.18}
\plotone{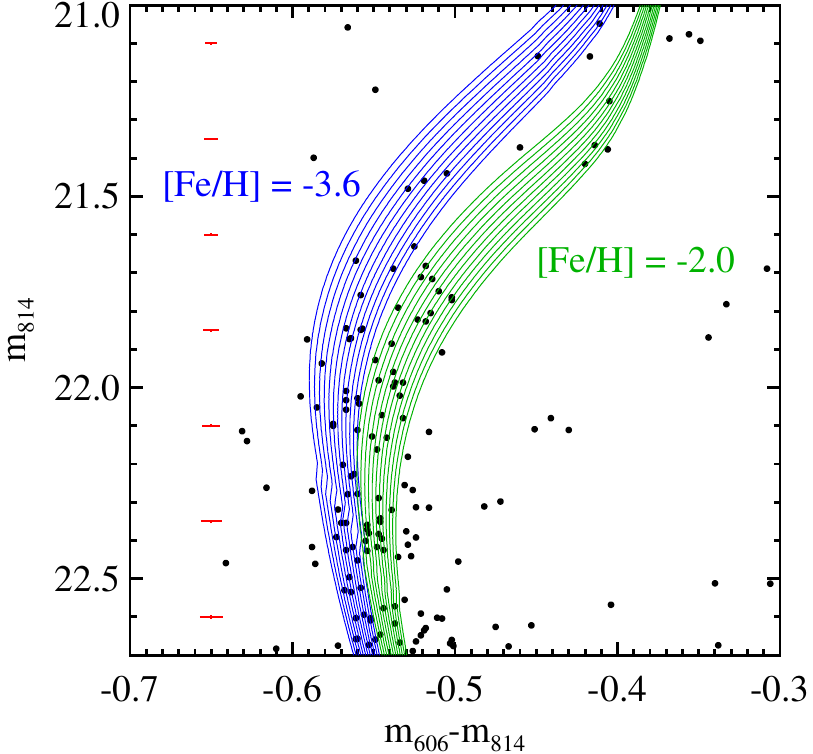}
\caption{Turnoff region of the Ret~II CMD in comparison to theoretical
  isochrones.  The range in age and metallicity covered by the full
  grid of isochrones is much larger than can be straightforwardly
  displayed in a single figure, so here we have selected metallicities
  representing the extremes of the Ret~II MDF.  The blue curves are
  Victoria-Regina isochrones at $\rm{[Fe/H]} = -3.6$ and the green
  curves are isochrones at $\rm{[Fe/H]} = -2.0$.  For each
  metallicity, 12 ages are shown in 200~Myr intervals from 12.1~Gyr
  (leftmost isochrone) to 14.3~Gyr (rightmost isochrone).  Typical
  photometric uncertainties are displayed in red along the left edge
  of the CMD. }
\label{fig:cmd_with_isochrones}
\end{figure}

Although we do find evidence for temporally extended star formation in
Ret~II, the available data do not enable us to select a single unique
combination of burst timing and duration.  To illustrate some of the
degeneracies in the fit results, in Fig.~\ref{fig:cmd_with_isochrones}
we show the MSTO region of the Ret~II CMD, accompanied by isochrones
spanning a range of ages for the most metal-poor and most metal-rich
stars in Ret~II.  The photometric uncertainties for individual stars
in this magnitude range correspond to age uncertainties (at constant
metallicity) of $\gtrsim500$~Myr and metallicity uncertainties (at
constant age) of $\gtrsim0.2$~dex.  With spectroscopic metallicity
measurements for individual MSTO stars (recall that the existing MDF
is determined entirely from brighter red giants), it may be possible
to derive improved constraints on the Ret~II SFH.

\begin{figure}
\epsscale{1.14}
\plotone{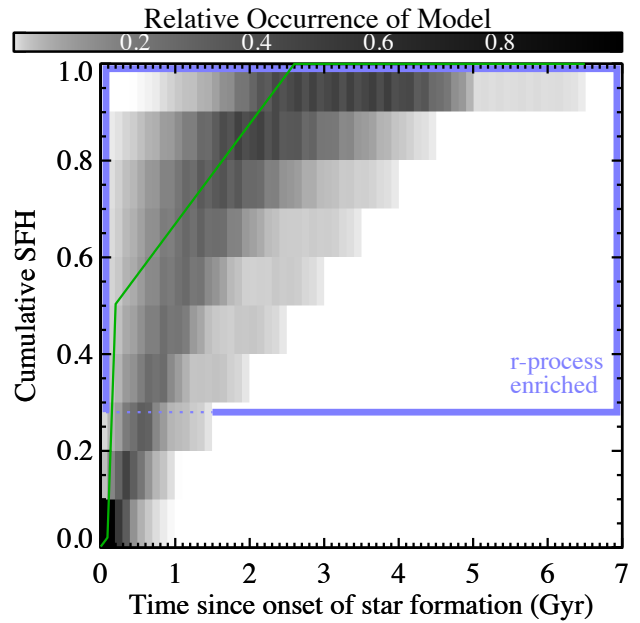}
\caption{Build-up of stars early in the history of Ret~II in the
  extended burst model.  The grayscale represents the density of
  models that pass through a given point, with black indicating 100\%
  of the models and white indicating 0\% (i.e., points that are not
  consistent with any of the models).  The models included in this
  figure are those with maximum likelihood scores within $1\sigma$ of
  the overall best fit.  Unlike the quasi-absolute ages shown in
  Fig.~\ref{fig:sfh}, in this plot the x-axis uses relative ages,
  where the beginning of star formation in each model is defined to
  occur at time~$=0$~Gyr. The green line displays the single best-fit
  model, which reaches 28\%\ of the stellar mass more quickly than the
  median of the acceptable models.  The blue outline indicates the
  portion of the parameter space that is observed to be enriched in
  \rproc\ elements.  The time at which a model crosses the boundary
  into the blue region therefore indicates the latest point at which
  the \rproc\ nucleosynthesis could have occurred.}
\label{fig:sfh_early}
\end{figure}

The early star formation history of Ret~II in the extended burst model
is illustrated in Fig.~\ref{fig:sfh_early}.  Here, we examine the
duration of star formation for all of the models with maximum
likelihood scores within $1\sigma$ of the best-fit model.  Note that
in this comparison, since the oldest burst of star formation does not
start at the same time in all models, we rely explicitly on relative
ages, normalized to the onset of star formation in each model.  These
results show that 28\% of the total stellar mass of Ret~II, matching
the fraction of non-\rproc-enhanced stars in the galaxy, had formed by
$500 \pm 200$~Myr after the system began to form stars.

We offer the results above with the caveat that the sample of stars
near the main sequence turnoff of Ret~II that are sensitive to the age
of the system is small.  With $M_{V} = -3.1 \pm 0.1$ \citep{mp18},
Ret~II is more than a factor of two less luminous than any of the UFDs
that we have previously analyzed \citep{brown12,brown14,simon21}, with
correspondingly fewer MSTO stars.  The sparsest CMD in that set of
galaxies belongs to Com ($M_{V} = -4.4$; \citealt{munoz18}), which had
275 MSTO stars in the \citet{brown14} data set.  The present Ret~II
observations include 176 stars along the main sequence and subgiant
branch between $m_{814} = 21.0$ and $m_{814} = 22.7$.  Although a
larger sample would clearly be beneficial, our ACS coverage already
includes essentially the entire area within the half-light radius of
the galaxy, and extends to $\sim1.4r_{\rm half}$ along the major axis
\citep{safarzadeh21}.  Assuming an exponential radial profile
\citep{mp18}, we estimate that the ACS mosaic includes 68\% of the
stars in Ret~II down to the magnitude limit of the data.  Thus, even
observing the entire area of Ret~II out to $>3 r_{\rm half}$ (as would
be straightforward with, e.g., the Roman Space Telescope;
\citealt{wang22}) would increase the number of Ret~II stars by no more
than $\sim50$\%, still resulting in a smaller sample of stars than was
obtained for Com.  Moreover, the number of foreground stars
contaminating the CMD would increase linearly with the observed area,
so the contamination of Ret~II by Milky Way stars would worsen with
wider coverage.

\section{ANALYSIS AND IMPLICATIONS}
\label{sec:discussion}

\subsection{Constraints on \rproc\ Nucleosynthesis}

In the largest spectroscopic study of Ret~II, $72^{+10}_{-12}$\% of
the stars in the galaxy were classified as \rproc-rich
\citep{ji22}.  Because the ejecta from the event that produced
the \rproc\ elements in Ret~II may have taken some time to uniformly
enrich the entire interstellar medium of the galaxy, this measurement
places an upper limit of $28^{+12}_{-10}$\% on the portion of
Ret~II stars that could have formed before the \rproc\ event.
In principle, if the enrichment was initially quite inhomogeneous and
the mixing time was long, the fraction of stars forming before the
event could have been considerably lower.

We therefore used the SFH of Ret~II from \S~\ref{sec:sfh_modeling} to
place a limit on when the \rproc\ enrichment must have occurred.  As
shown in Fig.~\ref{fig:sfh_early}, 28\% of the stars had formed within
$500 \pm 200$~Myr of the onset of star formation in Ret~II.  Thus, the
\rproc\ nucleosynthesis in Ret~II must have occurred no more than $500
\pm 200$~Myr after its first stars formed.  This result is consistent
with the star formation and chemical enrichment timescales seen in
hydrodynamic simulations of UFDs.  Specifically, \citet{tarumi20}
found that complete mixing of \rproc\ ejecta from a neutron star
merger occurs within 250~Myr, and \citet{jeon21} showed that
exclusively \rproc-rich stars are formed less than 100~Myr after the
merger event.  Although the present observational limit is not
strongly constraining in this context, it does demonstrate that long
delay times of $\gtrsim3$~Gyr, such as those inferred for the only
confirmed neutron star merger, GW170817
\citep[e.g.,][]{blanchard17,pan17}, are incompatible with the
enrichment of Ret~II.

The distribution of delay times between the formation of a binary
neutron star system and its eventual merger is currently poorly known
\citep[e.g.,][]{mv16,blanchard17,sb19,ss20}.  However, chemical
abundances in both the Milky Way and dwarf galaxies suggest that
\rproc\ nucleosynthesis (whether from merging neutron stars or not)
must occur rapidly in some cases
\citep[e.g.,][]{bp19,simonetti19,ss20}.  Galaxy formation simulations
also support rapid \rproc\ enrichment
\citep[e.g.,][]{vandevoort20,jeon21}.  Among the possible progenitor
systems that could lead to \rproc\ element production, rare
core-collapse SNe such as collapsars or magnetorotationally-driven SNe
would create \rproc\ material within $\sim10$~Myr, entirely consistent
with the SFH limits for Ret~II.  Assuming a \citet{salpeter55} initial
mass function for stars above 1~M$_{\odot}$, the total number of
core-collapse supernovae between 8 and 50~M$_{\odot}$ in Ret~II would
be $\sim180$, with $\sim20$ of these at masses above the
28~M$_{\odot}$ threshold suggested by \citet{taddia19} and
\citet{bm22} for some collapsars.  Rare core-collapse SNe are
therefore plausible in Ret~II from a stellar populations perspective
as well.  Neutron star mergers are also very likely to be compatible
with Ret~II, so long as the initial conditions of the binary neutron
stars allow them to merge within $\sim500$~Myr
\citep{bp19,safarzadeh19,andrews20}.  Despite this presumed
consistency with neutron star merger timescales, it is worth noting
that the observed lanthanide fraction for Ret~II stars is much higher
than that inferred for GW170817, so if Ret~II was enriched by a
neutron star merger then there must be a large range in lanthanide
fractions for different merger events \citep{jdh19}.

One additional scenario for the \rproc\ nucleosynthesis in Ret~II that
could be considered is that the \rproc\ elements were produced
directly by Population~III (Pop~III) stars
\citep[e.g.,][]{roederer14,mardini20}.  This connection between early
\rproc\ enrichment and the first stars might be expected if, for
example, collapsars are a major \rproc\ site and the Pop~III initial
mass function was top-heavy
\citep[e.g.,][]{bromm99,nu01,stacy16,stacy22}.  In that case, the
occurrence of a collapsar would be more likely, both because of the
increased number of massive stars per stellar mass formed and because
Pop~III stars likely have lower mass-loss rates, so that they can
retain the high masses and high angular momentum needed for collapsars
(as well as jet-driven SNe).  If no low-mass
($\lesssim0.8$~M$_{\odot}$) Pop~III stars were formed in Ret~II, then
the first generation of stars would have left behind only chemical
signatures, without contributing to the present-day stellar population
of the galaxy.  In this case, simulations suggest a delay of up to
$\sim100$~Myr before the formation of the first metal-enriched stars
\citep{magg22}, which would not be detectable as part of the overall
delay time between the production of \rproc\ material and the
formation of the \rproc-enhanced stars given the methodology we used
in \S~\ref{sec:sfh_modeling}.  However, this course of events would
make it difficult to explain the uniformity of the \rproc-enrichment
among the bulk of the Ret~II stars \citep{ji22}, which requires
complete mixing of the earlier nucleosynthetic products, as well as
the 28\% of the stars where neutron-capture species have not been
detected.  The very low \rproc-abundances in the latter set of stars
require either substantial inhomogeneities in the star-forming gas
within Ret~II, or perhaps that these stars were originally formed in a
different dwarf galaxy that did not feature a prolific \rproc\ event
and were later accreted by Ret~II.

\subsection{The Quenching of Reticulum~II}

As mentioned in \S~\ref{sec:sfh_modeling}, the SFH of Ret~II when fit
with the instantaneous burst model closely resembles those of the
lowest-luminosity members of the \citet{brown14} sample, Com ($M_{V} =
-4.4$), Leo~IV ($M_{V} = -5.0$), and CVn~II ($M_{V} = -5.2$).  These
galaxies each formed $\gtrsim80$\% of their stars in an initial burst
before reionization and have mean ages of $>13$~Gyr.  A small amount
of star formation as late as $z=2$ ($\sim10.5$~Gyr ago) cannot be
ruled out in any of these systems.  The Ret~II SFH shown in
Fig.~\ref{fig:sfh} is similar both qualitatively and quantitatively,
with a mean age of $13.7 \pm 0.2$~Gyr and more than 80\% of its
stellar mass in place at the earliest ages ($>12$~Gyr ago).  On the
other hand, our Ret~II models prefer not to have 100\%\ of the stars
forming by $z=6$, whereas that SFH is allowed for each of the other
UFDs listed above.

The Ret~II SFH is consistent with the general paradigm for quenching
in UFDs discussed in previous papers, where the large majority of the
star formation is complete before the end of reionization
\citep[e.g.,][]{brown14,wheeler15,rw19,applebaum21,sacchi21,simon21}.
Based on its Gaia DR2 proper motion, \citet{fillingham19} derived an
infall time for Ret~II of $10.2^{+1.1}_{-2.4}$~Gyr, which could be
consistent with the final cessation of star formation in Ret~II, but
occurred well after the star formation rate dropped to a small
fraction of its peak value.  We note that simulations show that star
formation in dwarf galaxies can continue at a low level for
$\sim1$~Gyr after reionization before the combination of heating and
lack of further accretion causes permanent quenching
\citep[e.g.,][]{onorbe15,rey19,wheeler19}.

A trickle of late-time star formation is also consistent with the
observed chemical evolution in Ret~II. \citet{ji22} found that the
most metal-rich star in Ret~II has a very low [Mg/Ca] ratio, out of 10
stars with Mg and Ca constraints. A low [Mg/Ca] ratio in dwarf
galaxies is often attributed to the integrated galactic initial mass
function \citep{weidner13, mcwilliam13}, where the low total gas mass
in a galaxy restricts the maximum mass of core-collapse SN progenitors
and thus the amount of Mg produced. In this case, the fraction of low
[Mg/Ca] stars would be expected to match the fraction of
post-reionization star formation in Ret~II, consistent with our
results.

Orbital studies including the gravitational potential of the LMC have
suggested that it has a significant gravitational influence on Ret~II,
with Ret~II classified as a recently-captured LMC satellite
\citep{patel20} or a longstanding member of the Magellanic group
\citep{battaglia22}.  In either case, the early environmental history
of Ret~II may be more difficult to untangle than previously assumed.
Nevertheless, the conclusion of \citet{rw19} that UFDs as a group
cannot be primarily quenched by environmental processes still holds,
and is perhaps strengthened by the addition of Ret~II to the set of
galaxies with well-determined SFHs.

\section{CONCLUSIONS}
\label{sec:conclusion}

We have derived the star formation history of the UFD Ret~II using
deep HST imaging covering most of the galaxy.  Similar to
previously-studied UFDs, we found that the galaxy is old, with most of
its stars forming shortly after the Big Bang.  A small minority of the
stars ($<15$\%) may have formed up to several Gyr later, at $z \sim
2$.

Although the SFH can be described by a model consisting of two
instantaneous bursts of star formation, we obtained slightly better
fits by allowing each burst to be extended in time.  With these
extended bursts, the best fit consisted of approximately half of the
stars forming in a short (100~Myr) burst and the other half forming in
a longer episode spanning 2.6~Gyr, both beginning at very early times.
In this scenario, a broad range of model parameters produce fits of
similar quality.  Recalling that 28\% of the stars in Ret~II are
lacking \rproc\ elements, we found that across the full set of models
consistent with the data, 28\% of the stars had been formed at
$t=500\pm200$~Myr after the beginning of star formation.  We therefore
concluded that the \rproc\ nucleosynthesis in the galaxy occurred no
later than $500\pm200$~Myr after the first Ret~II stars formed.  This
upper limit on the time delay between initial star formation and
production of \rproc\ material is consistent with either rare
core-collapse supernovae or a neutron star merger site for the \rproc,
with the constraint that the merger would need to occur relatively
quickly in the latter case.  \rproc\ sources with long delay times
($\gtrsim1$~Gyr) are ruled out in Ret~II.

The SFH of Ret~II shows a sharp decline around or before the time of
reionization, consistent with the possibility that photo-heating from
the increased ultraviolet radiation at that time was largely
responsible for quenching the galaxy.

Despite the increase in the number of UFDs with detailed chemical
abundance measurements in the last few years, the extreme
\rproc\ enrichment of Ret~II has remained unique.  The closest analog
is Tucana~III \citep{hansen17,marshall19}, but its \rproc\ abundances
are an order of magnitude lower and the classification of Tuc~III as a
dwarf galaxy has still not been confirmed \citep{simon17,baumgardt22}.
Identifying additional examples of this phenomenon, especially among
more luminous UFDs where the SFH can be determined more accurately,
would be helpful to improve the constraint on the source of the
\rproc\ elements.  In addition, it would be interesting to study the
\rproc\ abundances in more detail in UFDs that contain very little
\rproc\ material.  If the fraction of stars containing any
\rproc\ elements in those galaxies can be measured as it has been for
Ret~II, new constraints on the low-yield \rproc\ source could be
obtained as well.

\acknowledgments 
We thank the referee for suggestions that improved the presentation of
our results.  This publication is based upon work supported by Program
number HST-GO-14766, provided by NASA through a grant from the Space
Telescope Science Institute, which is operated by the Association of
Universities for Research in Astronomy, Incorporated, under NASA
contract NAS5-26555.  J.D.S. was also partially supported by the
National Science Foundation under grant AST-1714873.  B.M.P. was
supported by an NSF Astronomy and Astrophysics Postdoctoral Fellowship
under award AST-2001663.  A.F. acknowledges support from NSF grant
AST-1716251.

Funding for the DES Projects has been provided by the U.S. Department
of Energy, the U.S. National Science Foundation, the Ministry of
Science and Education of Spain, the Science and Technology Facilities
Council of the United Kingdom, the Higher Education Funding Council
for England, the National Center for Supercomputing Applications at
the University of Illinois at Urbana-Champaign, the Kavli Institute of
Cosmological Physics at the University of Chicago, the Center for
Cosmology and Astro-Particle Physics at the Ohio State University, the
Mitchell Institute for Fundamental Physics and Astronomy at Texas A\&M
University, Financiadora de Estudos e Projetos, Funda{\c c}{\~a}o
Carlos Chagas Filho de Amparo {\`a} Pesquisa do Estado do Rio de
Janeiro, Conselho Nacional de Desenvolvimento Cient{\'i}fico e
Tecnol{\'o}gico and the Minist{\'e}rio da Ci{\^e}ncia, Tecnologia e
Inova{\c c}{\~a}o, the Deutsche Forschungsgemeinschaft and the
Collaborating Institutions in the Dark Energy Survey.

The Collaborating Institutions are Argonne National Laboratory, the
University of California at Santa Cruz, the University of Cambridge,
Centro de Investigaciones Energ{\'e}ticas, Medioambientales y
Tecnol{\'o}gicas-Madrid, the University of Chicago, University College
London, the DES-Brazil Consortium, the University of Edinburgh, the
Eidgen{\"o}ssische Technische Hochschule (ETH) Z{\"u}rich, Fermi
National Accelerator Laboratory, the University of Illinois at
Urbana-Champaign, the Institut de Ci{\`e}ncies de l'Espai (IEEC/CSIC),
the Institut de F{\'i}sica d'Altes Energies, Lawrence Berkeley
National Laboratory, the Ludwig-Maximilians Universit{\"a}t
M{\"u}nchen and the associated Excellence Cluster Universe, the
University of Michigan, NSF's NOIRLab, the University of Nottingham,
The Ohio State University, the University of Pennsylvania, the
University of Portsmouth, SLAC National Accelerator Laboratory,
Stanford University, the University of Sussex, Texas A\&M University,
and the OzDES Membership Consortium.

Based in part on observations at Cerro Tololo Inter-American
Observatory at NSF's NOIRLab (NOIRLab Prop. ID 2012B-0001; PI:
J. Frieman), which is managed by the Association of Universities for
Research in Astronomy (AURA) under a cooperative agreement with the
National Science Foundation.

The DES data management system is supported by the National Science
Foundation under Grant Numbers AST-1138766 and AST-1536171.  The DES
participants from Spanish institutions are partially supported by
MICINN under grants ESP2017-89838, PGC2018-094773, PGC2018-102021,
SEV-2016-0588, SEV-2016-0597, and MDM-2015-0509, some of which include
ERDF funds from the European Union. IFAE is partially funded by the
CERCA program of the Generalitat de Catalunya.  Research leading to
these results has received funding from the European Research Council
under the European Union's Seventh Framework Program (FP7/2007-2013)
including ERC grant agreements 240672, 291329, and 306478.  We
acknowledge support from the Brazilian Instituto Nacional de Ci\^encia
e Tecnologia (INCT) do e-Universo (CNPq grant 465376/2014-2).

This manuscript has been authored by Fermi Research Alliance, LLC
under Contract No. DE-AC02-07CH11359 with the U.S. Department of
Energy, Office of Science, Office of High Energy Physics.

This research has made use of NASA's Astrophysics Data System
Bibliographic Services.

The HST data presented in this paper were obtained from the Mikulski
Archive for Space Telescopes (MAST) at the Space Telescope Science
Institute. The specific observations analyzed can be accessed via
\dataset[doi:10.17909/y62d-3794]{https://doi.org/10.17909/y62d-3794}.

\facility{HST (ACS)}

\software{DAOPHOT-II \citep{stetson87}, DOLPHOT
\citep{dolphin00}}

\bibliographystyle{apj}
\bibliography{main}{}

\clearpage

\end{document}